\documentclass[conference]{IEEEtran}
\IEEEoverridecommandlockouts
\usepackage{cite}
\usepackage{amsmath,amssymb,amsfonts}
\usepackage{algorithmic}
\usepackage{graphicx}
\usepackage{textcomp}
\usepackage{xcolor}

\usepackage{colortbl}
\usepackage{tabulary}
\usepackage{etoolbox}

\def\BibTeX{{\rm B\kern-.05em{\sc i\kern-.025em b}\kern-.08em
    T\kern-.1667em\lower.7ex\hbox{E}\kern-.125emX}}
\begin{document}

\title{Learning From Peers: A Survey of Perception and Utilization of Online Peer Support Among Informal Dementia Caregivers}

\author{Zhijun Yin$^{1,2}$, Lauren Stratton$^3$, Qingyuan Song$^2$, Congning Ni$^2$, Lijun Song$^2$, Patricia A. Commiskey$^1$, \\ Qingxia Chen$^1$, Monica Moreno$^3$, Sam Fazio$^3$, Bradley A. Malin$^{1,2}$\\ 
$^1$Vanderbilt University Medical Center, Nashville, USA\\
$^2$Vanderbilt University, Nashville, USA\\
$^3$Alzheimer's Association, Chicago, USA\\
\{zhijun.yin, cindy.chen, patricia.commiskey, b.malin\}@vumc.org\\
\{qingyuang.song, congning.ni, lijunsong\}@vanderbilt.edu\\
\{lstratton, mmoreno, sfazio\}@alz.org
}

\maketitle

\begin{abstract}
Informal dementia caregivers are those who care for a person living with dementia (PLWD) without receiving payment (e.g., family members, friends, or other unpaid caregivers). These informal caregivers are subject to substantial mental, physical, and financial burdens. Online communities enable these caregivers to exchange caregiving strategies and communicate experiences with other caregivers who they generally do not know in real life. Research has demonstrated the benefits of peer support in online communities, but they are limited in focusing merely on caregivers who are already online users. In this paper, we designed and administered a survey to investigate the perception and utilization of online peer support from 140 informal dementia caregivers (with 100 online-community caregivers). Our findings show that the behavior to access any online community is only significantly associated with their belief in the value of online peer support ({$\mathbf{p=0.006}$}). Moreover, 33  (83\%) of the 40 non-online-community caregivers had a belief score above 24, a score assigned when a neutral option is selected for each belief question. The reasons most articulated for not accessing any online community were no time to do so (14; 10\%), and insufficient online information searching skills (9; 6\%). Our findings suggest that online peer support is valuable, but practical strategies are needed to assist informal dementia caregivers who have limited time or searching skills.
\end{abstract}


\section{Introduction}
Alzheimer’s disease, the most common cause of dementia, is a brain disorder that affects the thinking, comprehension, and learning capacity of more than 6 million Americans and is the seventh leading cause of death in the US\cite{AFF2022}. An estimated 80\% of people living with Alzheimer’s disease or related dementia (PLWD) are cared for by unpaid informal caregivers (e.g., family members, friends, or other unpaid caregivers)\cite{corbett2012systematic}. In 2021, over 11 million informal dementia caregivers provided 16 billion hours of care to PLWD\cite{AFF2022}. While this care was valued at nearly \$271.6 billion, it imposed a substantial physical, financial, and mental burden on these informal caregivers\cite{oh2015motivations}. Additionally, 30\% of informal dementia caregivers are aged 65 or older\cite{AFF2022}. They are likely to experience reduced social engagement due to caring for PLWD, which increases their risk of developing Alzheimer's disease or some other dementia\cite{barnes2004social,boss2015loneliness,fratiglioni2000influence} and early death\cite{holt2010social,holt2015loneliness}. To ensure sufficient support for both informal caregivers and care recipients, it is essential for society to develop effective support mechanisms for the needs of informal dementia caregivers\cite{wennberg2015alzheimer}.

Research on how to best support informal dementia caregivers has focused primarily on assistance from credentialed professionals\cite{devor2008educational,kwok2014effectiveness,cheng2020benefit}. This type of assistance can improve a caregiver’s emotional well-being and caring skills, but maintaining this assistance over time can be difficult to achieve on a large scale. This is due, in part, to an insufficiently sized workforce, limited financial support
\cite{sanders2008experience,woods2016remcare}, the stigma of asking for help, and difficulties encountered when leaving individuals with dementia\cite{brodaty2005caregivers,ta2023scoping}. In addition, if they lack a shared experience, it may be difficult for healthcare professionals or other family members to respond to the specific needs of informal caregivers. This perception 
 that ``they simply do not understand” \cite{bruinsma2022they} can contribute to feelings of loneliness \cite{beeson2003loneliness}, which were found to be negatively associated with the health and well-being of these caregivers \cite{kovaleva2018chronic}. 

The integration of the Internet into daily life has enabled many people, including informal caregivers, to discuss health-related topics in online social media platforms\cite{yin2017talking,yin2017power}. For example, ALZConnected (https://www.alzconnected.org) organized by the Alzheimer's Association, is the largest online community for PLWD and their caregivers in North America. ALZConnected has accumulated tens of thousands of online users to discuss a broad range of topics regarding dementia caregiving and disease management \cite{cheng2022please}. Through online communities, informal dementia caregivers seek support and are willing to share experiences and practical information that they believe will assist other caregivers \cite{erdelez2019online}. A study analyzing an Alzheimer’s caregiver group on Facebook found that online peer social support had decreased the caregivers' burden while increasing their emotional and informational well-being\cite{oh2015motivations}. Similarly, a survey found that increased online activity among caregivers was associated with lower levels of depression and loneliness\cite{leszko2020role}. In addition, many online communities provide the added benefits like supporting anonymity, asynchronous participation, and connection to numerous caregivers without physical location and time constraints\cite{hodgkin2018emerging}, which provide a cost-effective and convenient way for informal dementia caregivers to gain support and access resources.

Based on the Internet utilization, informal dementia caregivers can be categorized into: 1) \textit{non-Internet caregivers} who never use the Internet; 2) \textit{non-online-community caregivers} who use the Internet but do not participate in online communities; and 3) \textit{online-community caregivers} who both use the Internet and participate in online communities. Current social media-based dementia caregiving research primarily focuses on online-community caregivers \cite{leszko2020role,hodgkin2018emerging,ni2022rough}. While improving the online experiences of these caregivers is important \cite{vaughan2018informal}, understanding how non-online-community caregivers perceive the value of online peer support is significant as well. This will inform the development of interventions for non-online-community caregivers in utilizing and benefiting from online peer support, thereby helping to mitigate the potential digital divide and decreasing existing health disparities in accessing online peer support \cite{olagoke2019digital}.


Hence, the primary objective of this research is to gain insight into the perceptions and utilization of online community support among informal dementia caregivers, specifically, non-online-community caregivers and online-community caregivers. To do so, we designed and administered a survey based on the Andersen and Newman Framework of Health Services Utilization (ANFHSU)\cite{andersen1995andersen}. ANFHSU is a classical model to identify and describe the factors that may affect a person's access to and utilization of health services. We distributed the survey through the Alzheimer's Association using their online website and ALZConnected community and obtained responses from 100 online-community caregivers and 40 non-online-community caregivers during a three-month period. Our findings suggest that online peer support is valuable, but practical strategies are needed to assist caregivers with limited time or online information-searching skills. This investigation marks the initial step towards addressing a long-term research objective, aiming to comprehensively elucidate the intricate mechanisms underlying online social support in dementia caregiving. 

\section{Methods}

We begin this section with the definition of online peer support. While peer support groups can be organized in an online format (e.g., via Zoom meetings), the online peer support in this paper refers to the communications between informal dementia caregivers who may not know each other in real life but connect in online communities, forums, or websites (e.g., Twitter, Facebook, Reddit, or ALZConnected). More broadly, since reading online caregiving discussions from other caregivers can be a way to learn information, resources, or caregiving skills, we also treat reading online posts as a behavior of seeking online peer support. This project was reviewed and deemed exempt from human subjects research by the Vanderbilt University Medical Center Institutional Review Board.

\subsection{Questionnaire Design}

\begin{figure*}[ht]
  \centering
    \includegraphics[width=0.8\textwidth]{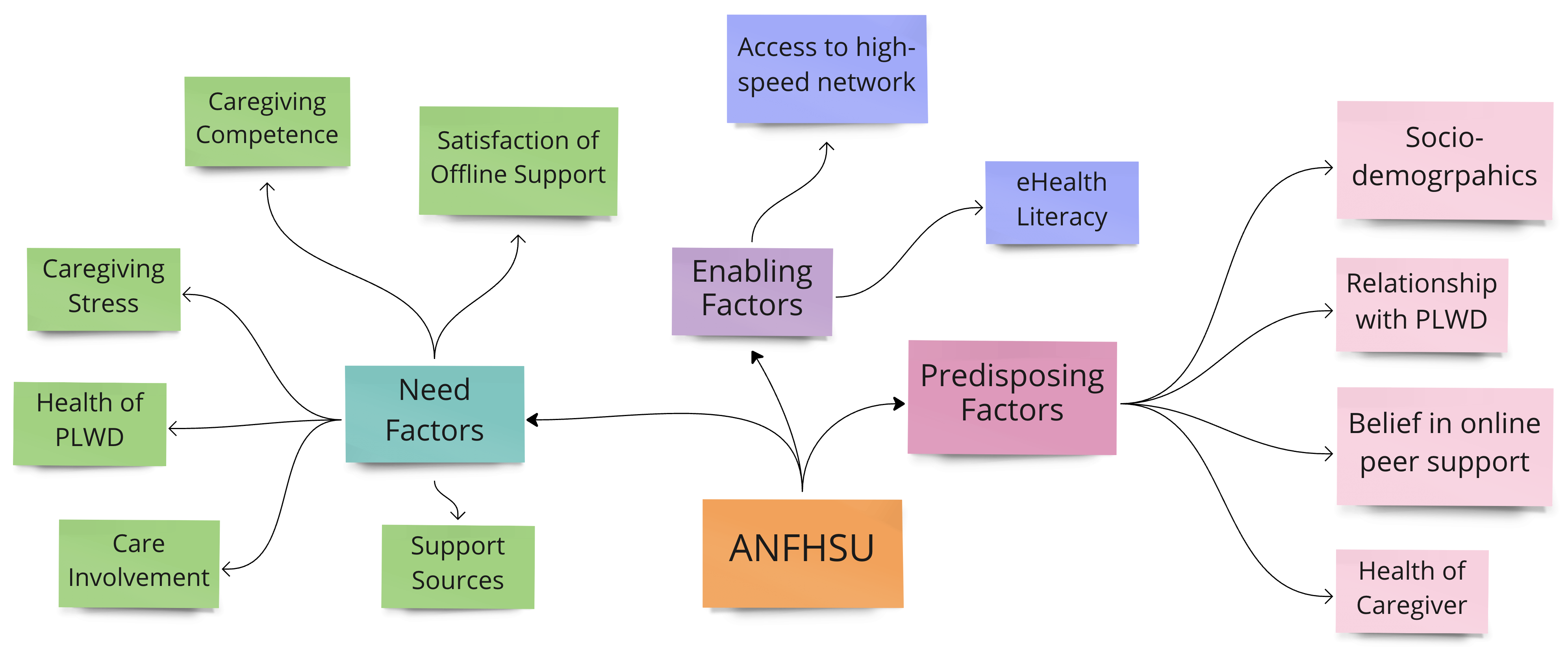}
 \caption{The \textit{Predispoing}, \textit{Enabling}, and \textit{Need} factors in the Andersen and Newman Framework of Health Service Utilization (ANFHSU). Please refer to the main text for the detailed components of each key factor.}
  \label{fig:ANFHSU}
\end{figure*}

We designed the survey based on the Andersen and Newman Framework of Health Services Utilization (ANFHSU), where an informal dementia caregiver’s access to or utilization of online peer support is considered to be a function of three characteristics (see Figure~\ref{fig:ANFHSU}): 

\textbf{Predisposing Factors}, which are the socio-cultural characteristics of individuals or other factors that exist before being a dementia caregiver. Specifically, we included the predisposing factors as follows:
    \begin{itemize}
        \item \textit{Socio-demographics}. They include age, biological sex, education level, race and ethnicity, occupational status, and income. We also inquired about the number of children that are taken care of (if applicable), since the ``sandwich generation'' caregivers encounter non-trivial life balance \cite{cheng2022please}.
        \item \textit{Relationship} between the caregiver and the PLWD. This is because caregivers who have different relationships with a PLWD face different caregiving challenges and burdens. 
        \item \textit{Belief} in the value of online peer support. We asked five 7-point Likert scale (from ``Strongly disagree'' to ``Strongly agree'') questions: \textit{Do you believe that reading online discussions from, or directly writing posts to discuss with, other caregivers whom you do not know in the real world will help a) find caregiving resources that you need, b) increase caregiving knowledge, c) increase understanding of the disease and patient, d) improve caregiving skills, e) increase confidence in caregiving, f) reduce caregiving stress or f) reduce the feeling of loneliness as a caregiver?}
        \item \textit{Health of the caregiver}. We collected the factor through a 7-point Likert scale question.
    \end{itemize}
    
\textbf{Enabling Factors}, which refer to the logistic aspects of accessing online peer support.
    \begin{itemize}
        \item \textit{Access} to high-speed network band, which is an important contributor to the digital\cite{sanders2021digital}. 
        \item \textit{eHealth Literacy}. It is measured through the eHealth Literacy Scale (eHEALS)\cite{norman2006eheals}, which evaluates if a caregiver has can process online health-related information. 
    \end{itemize}

\textbf{Need Factors}, which refer to the most immediate cause of seeking online peer support.
    \begin{itemize}
        \item  \textit{Health of the PLWD}. For simplicity, caregivers report the dementia stage for PLWD as an early, middle, or late stage.
        
        \item \textit{Care Involvement}. This is measured through two questions: 1) How long have you been taking care of the PLWD (\textit{caregiving duration})? and 2) how frequently do you care for the PLWD per week (\textit{caregiving workload})? This design is based on the fact that dementia caregiving is a long-term dynamic process, and the weekly caregiving workload will affect the need for online peer support.   
        \item \textit{Caregiving Challenges}. This question collects the most challenging issues that caregivers have faced when caring for the PLWD. 
        \item \textit{Offline Support Sources}. Where to obtain support in the real world to handle the aforementioned challenges?
        \item \textit{Satisfaction} with any support received in offline environments (7-point Likert scale). 
        
        \item \textit{Caregiving stress}. This is important psychological well-being that many interventions focus on improving. The intuition is that the more stress a caregiver experiences, the more likely they will seek support. We measure this factor using the Zarit Burden Interview (12-item) scale\cite{bedard2001zarit}. 
        \item \textit{Caregiver Competence}. This is a self-evaluation of one's capacity to care for the PLWD. We measure this factor using the CARERS Interview (4-item) scale\cite{pan2020examining}. 
    \end{itemize}
    The rationale of the design of the survey is that, if an informal dementia caregiver finds offline support insufficient in solving their caregiving challenges, they may turn to online environments for peer support.

In addition to the ANFHSU factors, we surveyed each participant’s experience in using online peer support in the past three months (\textit{behavior}). For the survey participants who are already online community users, we identified, \textit{1) Which online platforms have you visited in the past three months? 2) How frequently did you visit those online communities?
3) Will you intend to revisit these online communities in the next three months (\textit{intention})? 4) If the answer to question 3) is yes, what are your motivations for revisiting these online communities?} For participants who are not online community users, we asked them an open-ended question regarding why they did not seek online peer support in the past three months. 

\begin{table*}[hbt!]
\begin{center}
\caption{Summary of the socio-demographics and caregiving basics of the 140 survey participants.}
\begin{tabular}{lrlr}
\hline
\rowcolor[HTML]{656565} 
{\color[HTML]{FFFFFF} \textbf{Characteristic }} & {\color[HTML]{FFFFFF} \textbf{\begin{tabular}[c]{@{}c@{}} Participants (N = 140)\end{tabular}}} & {\color[HTML]{FFFFFF} \textbf{Characteristic}} & {\color[HTML]{FFFFFF} \textbf{\begin{tabular}[c]{@{}c@{}} Participants (N = 140)\end{tabular}}} \\ \hline
\rowcolor[HTML]{C0C0C0} 
\textbf{Age} & \textbf{} & \textbf{Age of PLWD} & \textbf{} \\ \hline
Mean (SD) & 54 ($\pm$ 13.5) & Mean (SD) & 76 ($\pm9.5$) \\
Range & [19, 87] & Range & [46, 97] \\ \hline
\rowcolor[HTML]{C0C0C0} 
\textbf{Gender} & \textbf{} & \textbf{Gender of PLWD} & \textbf{} \\ \hline
Female & 123 (87\%) & Female & 91 (65\%) \\
Male & 16 (11\%) & Male & 49 (35\%) \\ \cline{3-4} 
Undifferentiated & 1 (1\%) & \cellcolor[HTML]{C0C0C0}\textbf{\begin{tabular}[c]{@{}l@{}}Relationship \end{tabular}} & \cellcolor[HTML]{C0C0C0}\textbf{} \\ \hline
\cellcolor[HTML]{C0C0C0}\textbf{Race} & \cellcolor[HTML]{C0C0C0}\textbf{} & Adult child & 71 (51\%) \\ \cline{1-2}
White & 126 (90\%) & Spouse/Partner & 41 (29\%) \\
Asian & 8 (6\%) & \begin{tabular}[c]{@{}l@{}}Other Relative \end{tabular} & 19 (14\%) \\
Black or African American & 5 (4\%) & Grandchild & 6 (4\%) \\
Unknown & 1 (1\%) & Neighbor & 1 (1\%) \\ \cline{1-2}
\cellcolor[HTML]{C0C0C0}\textbf{Ethnicity} & \cellcolor[HTML]{C0C0C0}\textbf{} & Friend & 1 (1\%) \\ \cline{1-2}
Not Hispanic or Latino & 130 (93\%) & Other & 1 (1\%) \\ \cline{3-4} 
Hispanic or Latino & 10 (7\%) & \cellcolor[HTML]{C0C0C0}\textbf{\begin{tabular}[c]{@{}l@{}}Dementia Stage\end{tabular}} & \cellcolor[HTML]{C0C0C0}\textbf{} \\ \cline{3-4} \cellcolor[HTML]{C0C0C0}\textbf{Education Level} & \cellcolor[HTML]{C0C0C0}\textbf{}
 & Early-stage & 24 (17\%) \\
4-year college degree above & 57 (41\%) & Middle-stage & 86 (61\%) \\
4-year college graduate & 32 (23\%) & Late-stage & 30 (21\%) \\ \cline{3-4} 
Some college or 2-year degree & 40 (29\%)  & \cellcolor[HTML]{C0C0C0}\textbf{\begin{tabular}[c]{@{}l@{}} Caregiving Duration\end{tabular}} & \cellcolor[HTML]{C0C0C0}\textbf{} \\ \cline{3-4} 
High school or GED & 11 (8\%) & \textless 6 months & 12 (9\%) \\ \cline{1-2}
\cellcolor[HTML]{C0C0C0}\textbf{Employment Status} & \cellcolor[HTML]{C0C0C0}\textbf{} & 6 - 12 months & 13 (9\%) \\ \cline{1-2}
Full-time & 57 (41\%) & 1 - 2 years & 35 (25\%) \\
Retired & 40 (29\%) & 2 - 4 years & 35 (25\%) \\
Part-time & 22 (16\%) & \textgreater 4 years & 45 (32\%) \\ \cline{3-4} 
\begin{tabular}[c]{@{}l@{}}Unemployed \end{tabular} & 21 (15\%) & \cellcolor[HTML]{C0C0C0}\textbf{\begin{tabular}[c]{@{}l@{}} Caregiving Workload\end{tabular}} & \cellcolor[HTML]{C0C0C0}\textbf{} \\ \hline
\cellcolor[HTML]{C0C0C0}\textbf{\#Cared Children} & \cellcolor[HTML]{C0C0C0}\textbf{} & \textless 1 time a week & 6 (4\%) \\ \cline{1-2}
1 & 15 (11\%) & 1 - 2 times a week & 21 (15\%) \\
2 & 20 (14\%) & 3 - 6 times a week & 20 (14\%) \\
3 & 2 (1\%) & Daily & 93 (66\%) \\ \cline{3-4} 
\textgreater 3 & 3 (2\%) &  & \multicolumn{1}{l}{} \\
Does not apply & 100 (71\%) &  & \multicolumn{1}{l}{} \\ \cline{1-2}
\cellcolor[HTML]{C0C0C0}\textbf{Annual Income} & \cellcolor[HTML]{C0C0C0}\textbf{} &  & \multicolumn{1}{l}{} \\ \cline{1-2}
Less than \$25,000 & 24 (17\%) &  & \multicolumn{1}{l}{} \\
\$25,000 - \$49,999 & 30 (21\%) &  & \multicolumn{1}{l}{} \\
\$50,000 - \$74,999 & 26 (19\%) &  & \multicolumn{1}{l}{} \\
\$75,000 - \$99,999 & 21 (15\%) &  & \multicolumn{1}{l}{} \\
\$100,000 - \$149,999 & 28 (20\%) &  & \multicolumn{1}{l}{} \\
\$150,000 or more & 11 (8\%) &  & \multicolumn{1}{l}{} \\ \hline
\end{tabular}
\label{tab:summary}
\end{center}
\end{table*}

\subsection{Implementation and Dissemination}\label{implementation}

We implemented the survey questionnaire in REDCap (https://www.project-redcap.org/), a secure web application for building and managing online surveys. We distributed the survey link through two online locations. The first is ALZConnected where we posted the survey link at the top of the thread board in its two major caregiver forums: 1) \textit{Caregivers Forum} and 2) \textit{Spouse or Partner Caregiver Forum}. The second is the Alzheimer's Association website (https://www.alz.org/) where an advertisement for this survey was created to allow any person who visited the website would have a chance to access the survey link. Since this website is a popular information source for dementia patients and caregiving, this is where we expected to obtain survey responses from non-online-community caregivers. 

\subsection{Analysis}\label{Analysis}
We performed three types of analysis. First, we summarized the answers to the multiple-choice questions by illustrating the distribution of each selected choice. This was done to paint a broad picture of who the responded caregivers are, and their perception and utilization of online peer support.

Second, we summarized the answers to the open-ended questions through manual annotation. This is not a trivial task because the responses were filled with free text and there were no pre-defined categories before the annotation. To address these issues, two authors (CN, QS) read the responses and annotated the categories for each open-ended question independently. Next, both annotators compared and discussed their summarized categories to create the final categories with category names and definitions. Then, both independently modified their annotations with the agreed categories, again. For each response, we adopted a conservative approach and only reported the interaction of the categories summarized by the two annotators.

Finally, we fitted a logistic regression model (R v4.2.2) to analyze how the proposed ANFHSU factors were associated with utilizing online peer support. Due to the small sample size, we converted some categorical variables into binary ones. Specially, we encoded \textit{gender} as female/non-female, \textit{race} as white/non-white, \textit{education level} as a 4-year college degree or above/below 4-year college degree, \textit{access to high-speed network} as yes/no, \textit{relationship} as spouse/non-spouse, and all the other categorical variables (e.g., annual income, dementia stage, caregiving duration, and caregiving workload) as ordinal variables (e.g., numerical values with equal distance). A two-sided p-value smaller than 0.05 is considered to be statistically significant.

\section{Results}
We distributed the survey link on November 14, 2022, leaving it open for over three months until Feb 23, 2023. During this time period, we collected responses from 172 dementia caregivers. Among these participants, 140 (81.4\%) completed the entire survey.

\subsection{Characteristics of Caregivers and their PLWD}

Table~\ref{tab:summary} summarizes the socio-demographics of the dementia caregivers, their relationship with PLWD, care involvement, and the PLWD's dementia stage. These caregivers are aged 19-87 with a mean of 54, and a majority are female (88\%) or White (90\%). Over 60\% of these caregivers had a 4-year college or plus degree, compared to 36\% who had below a 4-year college education level. Only 40\% of these caregivers were employed, and 29\% took care of the PLWD and at least one child simultaneously. The caregivers' annual income was approximately uniformly distributed from less than \$25,000 to a range between \$100,000 and \$149,999. Only 8\% of these caregivers earned more than \$149,999. 

The age of the PLWDs ranged from 46-97 with a mean of 76, and 65\%  were female. A majority of the PLWD's caregivers were their adult children (51\%), followed by spouses or partners (29\%), and other relatives (14\%). Over 60\% of these PLWDs were at the middle stage, 21\% were at the late stage, and 17\% were at the early stage. 18\% of the respondents reported taking care of their PLWD for $<1$ year,  32\% for $>4$ years, and 66\% provided daily care.




\begin{figure*}[h]
  \centering
    \includegraphics[width=.8\textwidth]{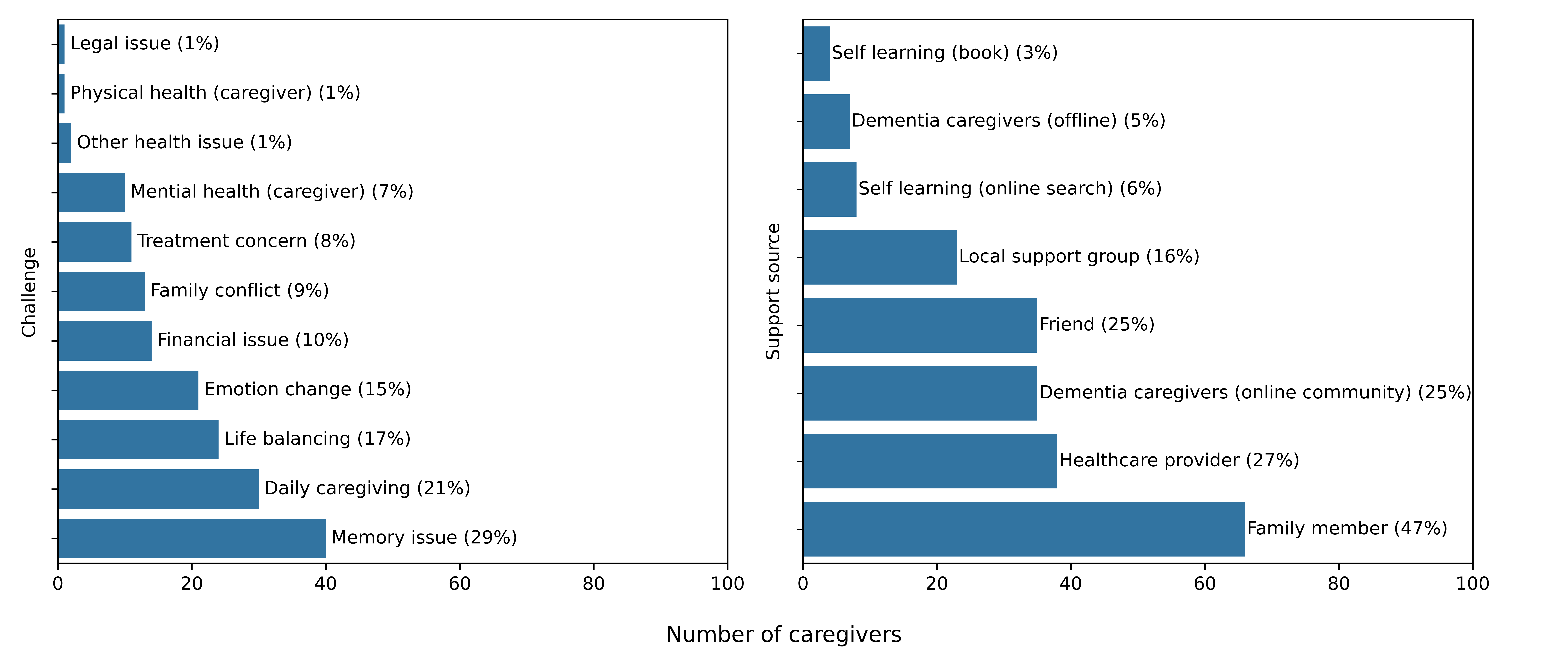}
 \caption{Summary of caregiving challenges (left) and specific support sources (right). }
  \label{fig:challenges}
\end{figure*}

\subsection{Caregiving Challenges and Support Sources}

Figure~\ref{fig:challenges} summarizes reported caregiving challenges and the sources where caregivers seek support. The main caregiving challenges include dealing with a PLWD's memory issue (29\%), supporting a PLWD in their daily life (such as showering and transportation; 21\%), and maintaining a balanced life (17\%). It should be noted that this category includes the ability to balance between 1) taking care of one's children and a PLWD; 2) work and caregiving; and 3) social life and caregiving. Dealing with the emotional fluctuations of the PLWD (15\%) and financial issues (10\%) were also major caregiving challenges. Some caregivers expressed concerns about the dementia treatment (8\%) for PLWD and their own mental health issues (7\%). 

\begin{figure*}[h]
  \centering
    \includegraphics[width=.8\textwidth]{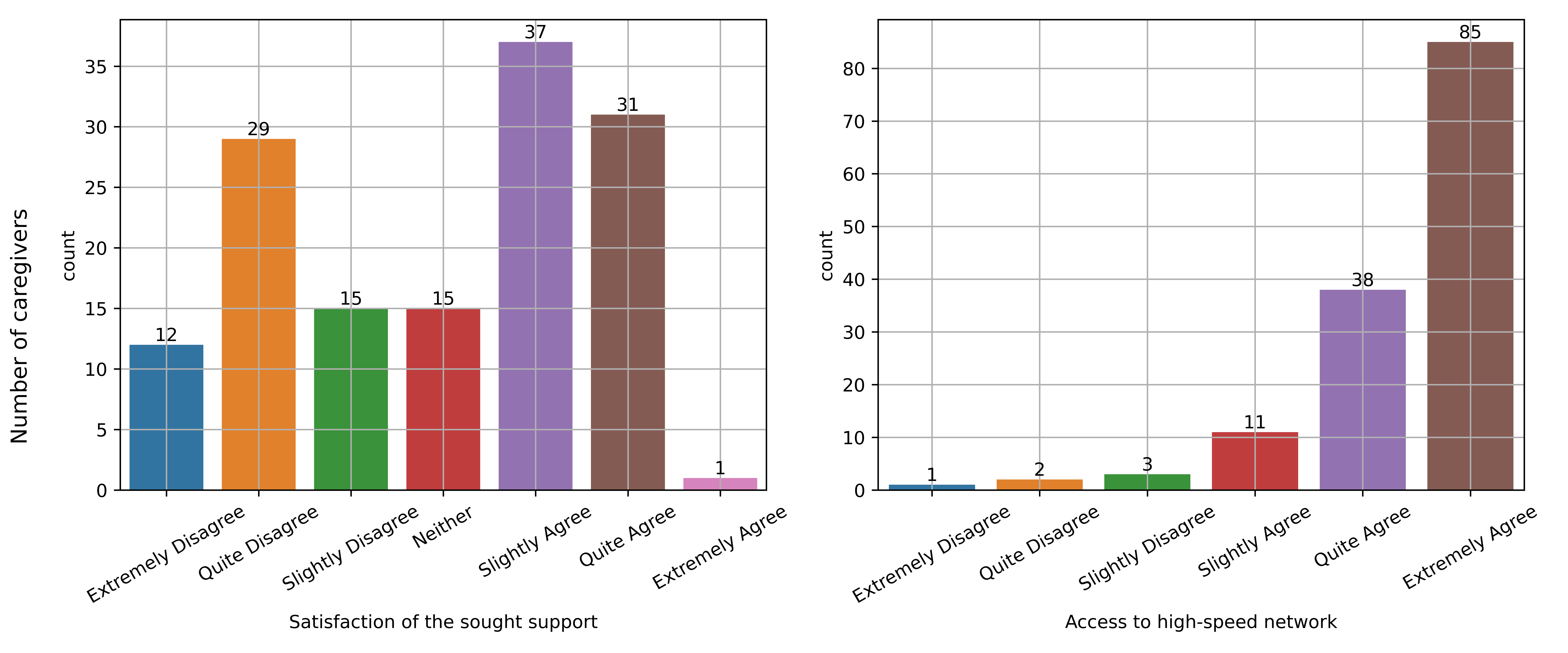}
 \caption{Distribution of satisfaction in support that was sought (left) and access to a high-speed network (right).}
  \label{fig:satisfaction}
\end{figure*}

\subsection{Satisfaction of Support and Network Access}

Figure~\ref{fig:satisfaction} depicts the distribution of caregivers’ satisfaction with the support they sought and their access to a high-speed network. 69 (49\%) caregivers reported an ``agree'' or above about the satisfaction of their sought support. Among the 54 caregivers who reported negative experience (39\%), 12 reported ``extremely disagree'' (9\%), indicating the challenging situation faced by these caregivers. Only 6 caregivers reported that they could not access a high-speed network.

\begin{figure*}[h]
  \centering
    \includegraphics[width=.8\textwidth]{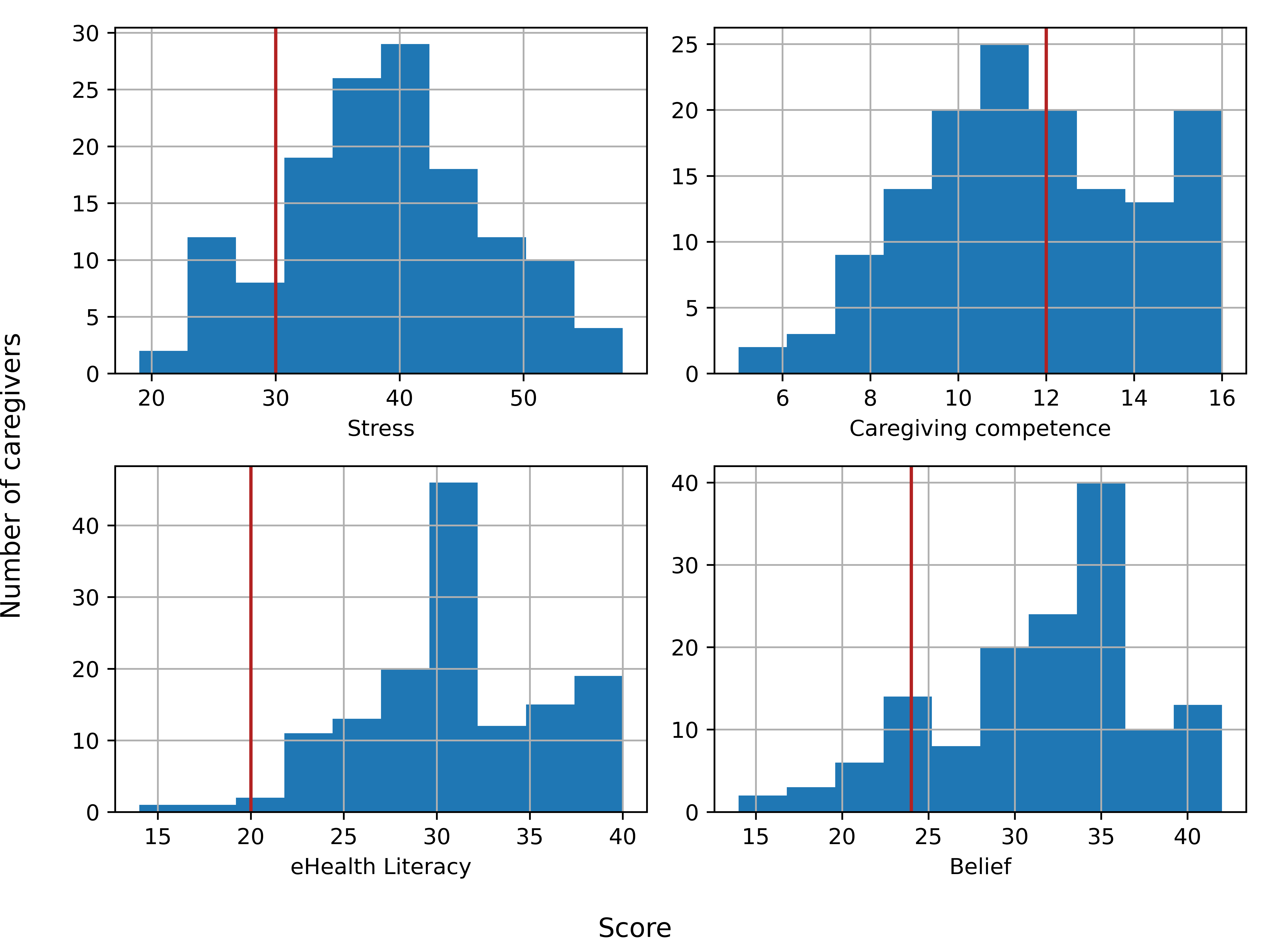}
 \caption{Distribution of scores of the four measures. The red vertical lines in caregiving \textit{Stress}, \textit{eHealth Literacy}, and \textit{Belief} in online peer support corresponds to the scores when ``neutral'' or ``undecided'' is selected for all the questions in each measure. The red vertical line in caregiving competence corresponds to the score when the ``fairly'' option is selected for all the questions.}
  \label{fig:scores}
\end{figure*}

\subsection{Measures of Caregiving Stress and Competence, eHEALS, and Belief}
Figure~\ref{fig:scores} illustrates the distribution of the score associated with caregiving stress, caregiving competence, eHealth literacy, and a caregiver's belief in the value of peer support from an online environment. The Chronbach’s Alpha, which measures the internal consistency of a questionnaire or survey (the higher, the better), of these measures was 0.86 (95\% confidence interval (CI) [0.82, 0.89]), 0.83 (95\% CI [0.78, 0.87]), 0.87 (95\% CI [0.84, 0.90]), and 0.87 (95\% CI [0.83, 0.89]), respectively. This indicates a very good internal consistency. Specifically, there are 121 (84\%) caregivers who had a stress score above 30 (the score when ``neutral'' is selected for all the questions, similar below), 67 (48\%) who had a competence score above 12, 138 (99\%) who had eHealth literacy score above 20, and 125 (89\%) who had belief score above 24. These results suggested that most of the caregivers were in stressful caregiving experiences, had high eHealth literacy, and believed in the value of online peer support. However, over half of these caregivers were not confident in their caregiving skills.

\begin{figure*}[h]
  \centering
    \includegraphics[width=.98\textwidth]{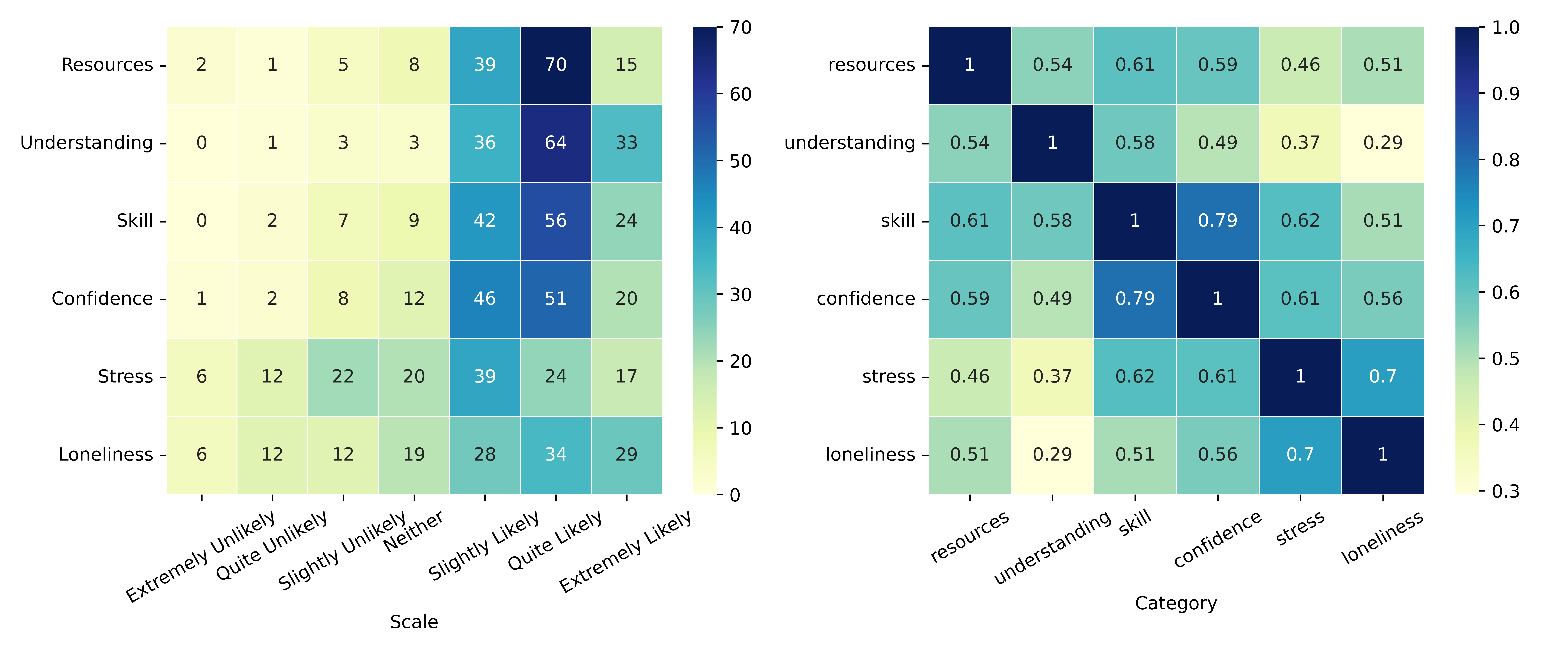}
 \caption{The number of survey participants in each pair of belief category and scale. \textit{Resources}: help find resources; \textit{Understanding}: understanding patients or the disease; \textit{Skill}: improve caregiving skill; \textit{Confidence}, increase caregiving confidence or competence; \textit{Stress}: reduce caregiving stress; \textit{Loneliness}: reduce the feeling of loneliness as a caregiver.}
 \label{fig:belief_score}
\end{figure*}

Figure~\ref{fig:belief_score} provides a detailed illustration, in the form of a heatmap, of the responses to each belief category (left) and the correlation in the responses (right). It shows that \textit{obtaining the needed resources}, \textit{increasing the understanding of the patient and disease}, and \textit{improving caregiving skills and confidence} were the top ``believed'' values of online peer support. By contrast, reducing \textit{caregiving stress} and \textit{loneliness} of being a caregiver was slightly more challenging. However, both categories received more ``Likely'' responses than ``Unlikely'' responses, with 57\% and 65\% of caregivers selecting at least ``Likely'' responses, respectively. There are several interesting observations here: First, reducing loneliness as a caregiver was highly correlated with reducing stress. Second, improving caregiving skills was highly correlated with increasing caregiving confidence. Third, reducing stress was correlated with improving caregiving skills and increasing caregiving confidence. While correlation does not necessarily imply causation, these results suggest that improving caregiving skills and reducing loneliness may help reduce caregiving stress. 

\subsection{Utilization of Online Peer Support}

Among the 140 survey participants, 40 caregivers (29\%) reported never using any online community to either read online caregiving discussions or discuss caregiving issues with other online peers in the past three months. The regression analysis showed that only the \textit{belief score} was statistically significantly associated with the utilization of online peer support ($\beta=0.11$, $sd=0.041$, $p=0.006$). All of the other factors, including socio-demographics, care duration and workload, dementia stage, access to a high-speed network, eHealth literacy, stress score, and competence score, did not have statistically significant effects. Particularly, 33 of the 40 (83\%) non-online-community caregivers had a belief score above 24, the score when ``neutral'' is selected for all the questions. Reasons for not using any online community to access online peer support were no time for online activities due to the intensive caregiving workload (14; 10\%), followed by limited online searching skills (9; 6\%), unreliable online information (6; 4\%), and security/privacy concerns (2; 1\%). Four (3\%) caregivers reported not wanting to spend time online after working on a computer during the daytime. Interestingly, 99 of the 100 online-community caregivers reported an intention to revisit online communities in the following three months. The only caregiver who did not do so said that too many sad stories in online communities made her worry about her father's future.  

\section{Discussion and Conclusion}\label{discussion}

This section discusses the implications of our research findings, limitations, and future directions. 

First, when comparing with the Alzheimer's Facts and Figures (AFF) in 2022\cite{AFF2022}, we observed that our survey recruited a similar proportion of caregivers who were Hispanic caregivers (7\% vs. 8\% in AFF) and Asian caregivers (6\% vs. 5\% in AFF). However, the proportion of Black or African American caregivers we recruited was far below that reported in AFF (4\% vs. 10\%), which suggests that Asian and Hispanic caregivers are more likely to participate in research studies compared to Black caregivers\cite{barrett2017engaging}. To gain greater insight into the situation, we reviewed the responses of the five caregivers who reported Black race. We found that two of the caregivers accessed online communities in the past three months, with one (age 49, college degree, stress score of 51) reporting this at least once a week and the other (age 50, more than a 4-year college degree, stress score of 19) reporting doing this when she had a caregiving question that needed an answer. Among the three remaining caregivers reporting Black race who did not access any online community in the past three months, one (age 60, 4-year college degree, stress score of 55) said ``\textit{I have used books and I think some information on the internet can be misinterpreted}'', and another one (age 29, some college or 2-year degree, stress score of 55) said ``\textit{In recently started looking for online communities}'', while the third (age 30, some college or 2-year degree, stress score of 52) said they were ``\textit{unaware}''. However, all their belief scores were above 24 and ranged from 25 to 42, which suggests that these three caregivers in our respondents believe in the value of online peer support, but may not know how to search online communities for peer support. Notably, the caregiver who relied on books for information seeking exhibited the lowest belief score of 25, indicating the survey result’s reliability.

Another main result of this study was that, whether an informal dementia caregiver accessed online health communities in the past three months depended upon their beliefs in the value of peer support obtained from online health communities but not upon their socio-demographics or any other ANFHSU factors. For example, one caregiver who said "\textit{I don't find comfort from strangers on the internet. I would love to and I am willing to go to an in-person meeting of a support group, but I don't have anyone to watch my mother so that I could attend}" exhibited a belief score of 14, which is logical based on their reason for not accessing any online communities. At the same time, it further highlights the dilemma that there is limited time to attend local support groups because of intensive caregiving \cite{brodaty2005caregivers}. Another caregiver with a belief score of 14 indicated the health insurance could only cover 40 days of at-home-nursing support in one year, which made their family feel so ``\textit{helpless and alone}''. Despite various reasons for not using online health communities in the past three months, over 80\% of respondents exhibited a belief score above 24. This suggests that online peer support is valuable to them, but an effective strategy to ``bridge'' their needs and the desired online peer support is needed. 

There are, however, several limitations to this study that can serve as a basis for future research. First, since we distributed our survey link in the ALZConnected online community and the Alzheimer’s Association’s website, our results may be biased towards online-community users. A less biased approach may be designed to collect data that reflects the perception and utilization of online peer support in the dementia caregiver population. Making online peer support beneficial to non-Internet caregivers is equally essential, but how to address the Internet access issue is a priority and is  beyond the scope of this study. Second, only a small percentage of the participants are caregivers reported Black race. Increasing participation with this group would increase the understanding of their perception and utilization of online peer support. Third, the analysis was based on 140 completed responses with limited power. While statistically significant findings were discovered, larger sample sizes are needed to investigate this research more accurately. Fourth, we asked what kinds of offline support were received to address the caregiving challenges in an open-ended question. Some caregivers included online peer support in their free text, such that the answer to the following question (satisfaction of the offline support) was not aligned with the original design. Future investigations should clarify or convert this question into a multiple-choice version. Finally, it is essential to study how to help caregivers without time or sufficient information-searching skills screen the online caregiving discussions they need. Overall, our findings serve as valuable evidence for leveraging online peer support to assist informal dementia caregivers effectively. 


\bibliographystyle{IEEEtran}
\bibliography{output}





\vspace{12pt}

\end{document}